\documentclass[aps,prb,reprint,superscriptaddress]{revtex4-2}

\usepackage{graphicx}
\usepackage{subcaption}
\usepackage{amsmath}

\begin{document}

\title{Spectral Emission Properties of a Nitrogen-doped Diamond(001) Photocathode: Hot Electron Transport and Transverse Momentum Filtering}

\author{Louis A. Angeloni}
\affiliation{Department of Physics, University of Illinois at Chicago, Chicago, Illinois 60607, USA}
\author{Sergey V. Baryshev}
\affiliation{Department of Electrical Engineering, Michigan State University, East Lansing, Michigan 48824, USA }
\author{Matthias M{\"u}hle}
\affiliation{The Fraunhofer USA, Inc. Center Midwest, East Lansing Michigan 48824, USA}
\author{W. Andreas Schroeder}
\affiliation{Department of Physics, University of Illinois at Chicago, Chicago, Illinois 60607, USA}

\date{\today}

\begin{abstract}
The electron emission properties of a single-crystal nitrogen-doped diamond(001) photocathode inserted in a 10kV DC photoelectron gun are determined using a tunable (235-410nm) ultraviolet laser radiation source for photoemission from both the back nitrogen-doped substrate face and the front homo-epitaxially grown and undoped diamond crystal face.  The measured spectral trends of the mean transverse energy (MTE) and quantum efficiency (QE) of the emitted electrons are both anomalous and non-monotonic, but are shown to be consistent with (i) the known physics of electron photoexcitation from the nitrogen substitution states into the conduction bands of diamond, (ii) the energy position and dispersion characteristics of the conduction bands of diamond in the (001) emission direction, (iii) the effective electron affinity of the crystal faces, (iv) the strong electron-(optical)phonon coupling in diamond, and (v) the associated hot electron transport dynamics under energy equipartition with the optical phonons.  Notably, the observed hot electron emission is shown to be restricted parallel to the photocathode surface by the low transverse effective masses of the emitting band states – a transverse momentum filtering effect.  
\end{abstract}

\maketitle

\section{Introduction}
	The generation of bright electron pulses with a low intrinsic transverse emittance (i.e., beam divergence) by photoemission from solid-state photocathodes are key to the success of today's and future scientific research \cite{1,2}; in particular, for x-ray free electron lasers (XFELs)  \cite{3,4}, ultrafast electron diffraction (UED) systems  \cite{5,6,7}, and (nanosecond to sub-picosecond) dynamic transmission electron microscopes (DTEMs)  \cite{8,9,10} employed for single-shot visualization of atomic-scale processes on fast  timescales  \cite{5,6,10,11,12,13,14,15,16}.  Indeed, even a modest factor of 2-3 reduction in the injected electron beam emittance is expected to enhance the performance (x-ray beam brilliance or emitted photon energy) of XFELs by an order of magnitude  \cite{17,18}. 

	Recent work has shown that the mean transverse energy (MTE) of the electrons emitted from solid-state photocathodes, while dependent upon temperature  \cite{19,20,21}, is determined primarily by the dispersion of the bulk energy bands and other electronic states from which they originate  \cite{22,23,24} and is limited by the physics of the recipient vacuum states  \cite{25}.  This has led to searches  \cite{26,27} for candidate oriented single-crystal photocathodes that exhibit an increase of their MTE with photon energy $\hbar\omega$ that is less than the ‘canonical’ monotonic $MTE = \frac{1}{3}\Delta E$ variation  \cite{1,28}, where $\Delta E = \hbar\omega-\phi$ is the excess photoemission energy for a photocathode with surface work function $\phi$. 

	The MTE of photo-emitted electrons, defined as $\frac{(\Delta p_{T})^2}{2m_0}$ where $\Delta p_{T}$ is their two-dimensional rms transverse momentum (i.e., parallel to the photocathode face) and $m_0$ is the free electron mass, will be restricted if the conserved transverse momentum of the emitting state in photoemission  \cite{29} is less than $\sqrt{2m_0\Delta E}$  \cite{22,25}.  In particular, if the emitting band states have a parabolic transverse dispersion of the form $\frac{p_T^2}{2m_T^*}$, where $m_T^*$ is the transverse effective mass, then the MTE for ‘thermionic’ emission from a Boltzmann distribution with an electron temperature $T_e$ and when $m_T^*$ is less than $m_0$ is given approximately by $MTE\approx\frac{m_T^*}{m_0}k_BT_e$ where $k_B$ is Boltzmann’s constant  \cite{25}.  The exploitation of such an emitting band structure dependence could then, of course, also reduce the rms normalized transverse emittance $\epsilon_n$ of the generated electron beam (or pulse) since, for a given rms source size $\Delta x$, $\epsilon_n = \frac{\Delta x\Delta p_x}{m_0c}$ where the one-dimensional rms transverse momentum $\Delta p_x = \frac{1}{\sqrt{2}}\Delta p_T$ for a cylindrically symmetric beam (about its direction of propagation) and c is the speed of light in a vacuum  \cite{28}. 

	Although not generally considered a good photocathode material due primarily to its large $\sim$6eV work function (valence band maximum to vacuum energy level)  \cite{30,31}, diamond has been shown to exhibit a striking 1eV negative electron affinity (NEA) on both (001)  \cite{32} and (111)  \cite{33} hydrogen-terminated crystal faces and, when combined with deep donor doping (e.g., nitrogen), photoemission at visible wavelengths has been observed from poly(nano)crystalline diamond films on a Mo substrate  \cite{34}.  From a more fundamental point of view, diamond is also an excellent and robust crystal material for understanding the influence of electronic band structure on the emission properties of solid-state photocathodes.  Of particular interest is that accurate modern Ab initio band structure calculations  \cite{35} show that the conduction band minimum (CBM) consists of six side valleys aligned in the $\Gamma$-X direction with an effective mass of 0.22$m_0$ transverse to the primary crystal directions (e.g. (001)) and an upper conduction band centered on the $\Gamma$ point with a relatively isotropic effective mass of about 0.4$m_0$.  Nitrogen doping, which generates two known dopant ‘defect’ energy levels at 2.4 and 4.7eV below the indirect conduction band minimum (CBM)  \cite{36,37}, allows these conduction band states to be populated by photo-excited electrons using blue to ultraviolet (UV) radiation ($\sim$3.0 to 5.5eV photon energies) accessible with today’s laser technology, thereby providing a means to study the emission properties of this promising photocathode material.

	In this paper, we characterize the spectral photoemission properties (MTE and quantum efficiency (QE)) of a diamond(001) photocathode homoepitaxially grown on a nitrogen-doped substrate.  The properties of electron emission from both the n-type substrate and undoped crystal faces are measured – the latter therefore requiring electron transport through the undoped region.  The obtained experimental data displays regions with a non-monotonic and anomalous spectral MTE dependence similar to that observed for nitrogen-doped ultra-nanocrystalline diamond  \cite{38} and polycrystalline Cs-Te  \cite{39} photocathodes; that is, a spectral region where the MTE decreases, rather than increases, with increasing photon energy.  It also displays clear evidence for the restrictive transverse effective mass dependence of the MTE for the observed predominantly sub-threshold photoexcited thermionic emission of hot electrons from diamond’s conduction band states – an effective transverse momentum filtering in the electron emission.  Further, the experimental data is shown to be consistent with the strong electron-optical phonon scattering dynamics in diamond  \cite{40} coupled with its high 160meV optical phonon energy  \cite{31,41}.
\section{The Nitrogen-doped diamond(001) photocathode}
	The studied nitrogen-doped diamond(001) photocathode sample was grown by microwave plasma assisted chemical vapor deposition (CVD) in a 2.45GHz reactor system  \cite{42,43}.  Briefly, in the reactor B configuration  \cite{43}, a recessed pocket holder was used for improved single-crystal diamond growth  \cite{42}.  A hydrogen plasma was struck and sustained while the CVD reactor was brought to a process pressure of 240Torr, at which point 5\% methane (the carbon source) was added to the gas composition.  The absorbed microwave power level was adjusted between 1800 and 1950W to achieve a constant substrate temperature of 900$^\circ$C for the crystal growth.  The resulting $\sim$4x4mm$^2$ sample consists of a ~0.5mm-thick single crystal and unintentionally-doped (possibly nitrogen doping less than 200ppb (parts per billion)  \cite{42}) diamond layer homoepitaxially-grown atop a ~400$\mu$m-thick, nitrogen-doped, type 2c, (001)-oriented diamond substrate.  The substrate side (back side) of the photocathode appears dark grey due to its 100-200ppm (parts per million) nitrogen doping and, although mostly flat, has a stepped surface with an irregular periodicity slightly less than the ~100$\mu$m incident laser beam size employed in the spectral characterization measurements.  In contrast, the growth side (front side) is optically flat and essentially transparent down to a wavelength of ~230nm corresponding to the fundamental 5.4eV indirect band gap of diamond  \cite{44}.  The photoelectron emission characteristics of this single-crystal photocathode were investigated for two geometries (accessed by physically turning over the sample) associated with electron emission from the same crystal surface upon which the exciting UV photons are incident; that is, (i) front-side illumination incident on the ‘undoped’ grown layer and (ii) back-side illumination on the nitrogen-doped diamond substrate.  Consequently, electron emission in the former case requires electron transport through the ~0.5mm-thick epitaxially-grown diamond as the undoped layer cannot be a source of electrons for the employed 3.0-5.3eV (230-410nm) tunable UV radiation source  \cite{23}.

	No surface preparation, such as hydrogen surface termination  \cite{31,32}, was performed on either face of the nitrogen-doped diamond(001) sample to affect the photocathode’s work function prior to the spectral measurement of its electron emission properties.  Consequently, the electron affinity $\chi$ – the difference between the energy of the conduction band minimum (CBM) and the vacuum level (see Figure 1) – of both the doped substrate and undoped (001) surfaces is expected to be positive  \cite{30,45}.  Specifically, the vacuum level on the undoped diamond(001) face should be close to where the lower conduction band intersects the X point of the Brillouin zone; that is, approximately 0.5eV above the CBM  \cite{30}.  Upward band bending, due to the n-type nitrogen doping  \cite{45}, should increase the effective value of $\chi$ to around 1eV for the doped (001) surface, assuming that the Fermi level is pinned close to the midpoint of the band gap at the crystal surface.  Clearly, a difference in the electron affinity of the two faces will affect electron emission since electrons in the vicinity of the CBM will see a potential step proportional to $\chi$ for transmission into the vacuum.
\section{Experimental methods}
	The spectral electron emission characteristics of the nitrogen-doped diamond(001) photocathode were determined using a 235-410nm tunable, p-polarized, sub-picosecond UV radiation source incident at 60$^\circ$ on the photocathode mounted in a DC gun based on the design analyzed in Ref. 46.  The laser-based tunable UV radiation source has been described elsewhere  \cite{23}.  Briefly, a front-end, mode-locked, 28MHz repetition rate, diode-pumped Yb:KGW laser oscillator  \cite{47} generating 250fs duration pulses at 1047nm is frequency doubled to pump a optical parametric amplifier (OPA) seeded by a continuum produced using a nonlinear photonic crystal fiber.  The amplified signal and idler radiation from the OPA is then sum frequency mixed in $\beta$-Barium borate (BBO) nonlinear crystals with the second and third harmonics of the Yb:KGW laser to produce the tunable UV radiation.  This radiation, with a beam diameter greater than 300$\mu$m, is incident on a 200$\mu$m-diameter tungsten pinhole which is relay-imaged using achromatic Al focusing mirrors with a de-magnification factor of 0.76 onto the photocathode surface, resulting in an ellipsoidal, near top-hat, incident laser beam with major and minor axes of 152 and 76$\mu$m, respectively.  The resultant enhanced beam pointing stability and constant laser beam size incident on the photocathode surface reduces statistical uncertainty in the extracted values of the MTE at the expense of only a modest deterioration in UV power stability due to inherent beam pointing instabilities of the radiation source. 

	The electrons photoemitted from the nitrogen-doped diamond(001) photocathode accelerated in the 10mm cathode-anode gap of the DC gun propagate over a distance of 45cm from the 5mm-diameter aperture gun anode to a 18mm-diameter micro-channel plate (MCP) detector (BOS-18 from Beam Imaging Solutions Inc.) incorporating two 10$\mu$m-pore chevroned plates with a total voltage-controlled avalanche gain of up to $\sim$10$^6$.  The amplified electron signal emerging from the chevroned plates is accelerated to typically 2keV over a short (few millimeters) distance before impinging on a P-43 phosphor screen to give an image resolution (point spread function width) of $\sim$25$\mu$m.  The optical signal from amplified electron beam on the phosphor screen is then recorded using 8:5 reduction relay imaging with two visible achromatic doublet lenses onto a CCD camera with 5$\mu$m pixel resolution.

	A detailed simulation of the photoemitted electron trajectories in the DC gun, using its numerically evaluated three-dimensional electric field distribution, and subsequent free-space propagation to the MCP detector is used to calibrate the initial photoemitted transverse electron momentum per pixel on the CCD camera image from a point source on the photocathode surface.  This simulation indicates that each 5$\mu$m CCD camera pixel is equivalent to an emitted transverse electron momentum of 0.002 $(m_0.eV)^{1/2}$ for a 10kV DC gun voltage.  Inclusion in the simulation of the photoemitted electron transverse momentum distribution at each UV photon energy (i.e. excess photoemission energy, $\Delta E$) then gives the expected measured electron beam shape, provided appropriate convolutions with both the incident UV laser beam size on the photocathode surface and the optically imaged point spread function of the MCP detection system are performed.  In practice, these convolution factors, which give the measurement system an inherent mean transverse energy (MTE) resolution of $\sim$1meV for a 10kV DC gun voltage, only start to affect the beam size measurements for MTE values less than 10meV; that is, become significant compared to the overall 5-10\% experimental measurement uncertainty.  As a result, the transverse momentum calibration, which has been verified using the known and benchmarked against theory photoemission from a Rh(110) photocathode  \cite{23}, is used in this work to directly determine the MTE of photoemitted electrons from the measured pixelated spatial electron beam signal.  In agreement with experimental observations, the electron trajectory simulation also indicates that the detected beam size and shape (and hence extracted MTE value) are not strongly perturbed by errors in positioning of the incident laser beam on the photocathode surface, as long as it is placed within $\pm$0.5mm of the DC electron gun’s central ‘optical axis’.

	The spectral dependance of the photoemission quantum efficiency (QE) is determined by integration over the detected electron beam signal and a measurement of the incident tunable UV laser power at the UV grade fused silica entrance window to the vacuum chamber.  This evaluation employs a calibration of the MCP detector for each set of applied plate and phosphor screen acceleration voltages with the known spectral QE dependance of the Rh(110) photocathode  \cite{23}.  Corrections due to the reflection losses at the fused silica window and at the surface of the nitrogen-doped diamond photocathode (2-3\% over the entire studied UV range for the incident p-polarized radiation,) then allow the QE per photon absorbed by the material to be evaluated.  As the wavelength dependance of the refractive index for fused silica  \cite{48} and diamond  \cite{49} are well known, the primary uncertainties in the QE determination are from the calibration of the MCP detector and the stability of the tunable UV radiation source over the typical 1-30s measurement duration.  While the former is partially off-set by the good agreement of the spectral QE trend for Rh(110) with our one-step photoemission simulation  \cite{23,25}, the latter can be as large as $\pm$20\% rms due to the cascade of nonlinear optical techniques used to generate the tunable UV radiation  \cite{23}.  As a result, rms uncertainties in the extracted QE values can reach $\pm$25\% – the value conservatively used in this paper.  Due to the enhanced sensitivity provided by the high avalanche gain in the MCP detector, QE values down to 10$^{-9}$ can be measured by the experimental system for incident UV pulse energies above $\sim$1pJ (i.e., above $\sim$100 photons/pulse), albeit at the expense of increased noise in the electron detection at the required longer CCD camera exposure times.
\section{Spectral MTE and QE measurements}
	The measured spectral emission properties of the nitrogen-doped diamond(001) photocathode are associated with the combined effects of (i) the energetics of electron photoexcitation from the nitrogen substitution (NS) states into the conduction bands of diamond  \cite{36,37} (see Figure 1), (ii) the strong electron-(optical)phonon coupling in diamond  \cite{40}, (iii) the position of the vacuum energy with respect to the bottom of the conduction band (i.e., the electron affinity and surface band bending due to the nitrogen doping)  \cite{30,32,45,50}, (iv) the position and dispersion in momentum space of the diamond conduction bands  \cite{35}, and (v) hot electron transport dynamics.  

	For the 100-200ppm nitrogen doping level in the substrate of the diamond photocathode, it is known that the mid-gap (i.e., UV-visible) absorption is dominated by two neutral substitution states; one, the upper nitrogen state, located at $\sim$2.4eV below the conduction band minimum (CBM) and the other, the lower nitrogen state, located $\sim$4.7eV below the CBM (i.e., $\sim$0.7eV above the valence band maximum)  \cite{36}.  At this doping level, measurements using the constant-photocurrent method on polycrystalline diamond grown by CVD  \cite{36} indicate that the upper nitrogen dopant state is quite broad with an energy half-width 1/e maximum (HW1/eM) of $\sim$0.6eV and so extends from $\sim$1.8 to $\sim$3.0eV below the CBM.  Even at significantly lower nitrogen dopant densities of $<$10ppm, the energy HW1/eM of this upper state is quite large $\sim$0.2eV  \cite{37} due to the Jahn-Teller effect.  On the other hand, the lower state has an energy HW1/eM of about 0.3eV but, importantly, has a dopant density that is an order of magnitude larger than that of the upper nitrogen substitution state  \cite{36}.
\begin{figure}
	   \includegraphics[width=\linewidth]{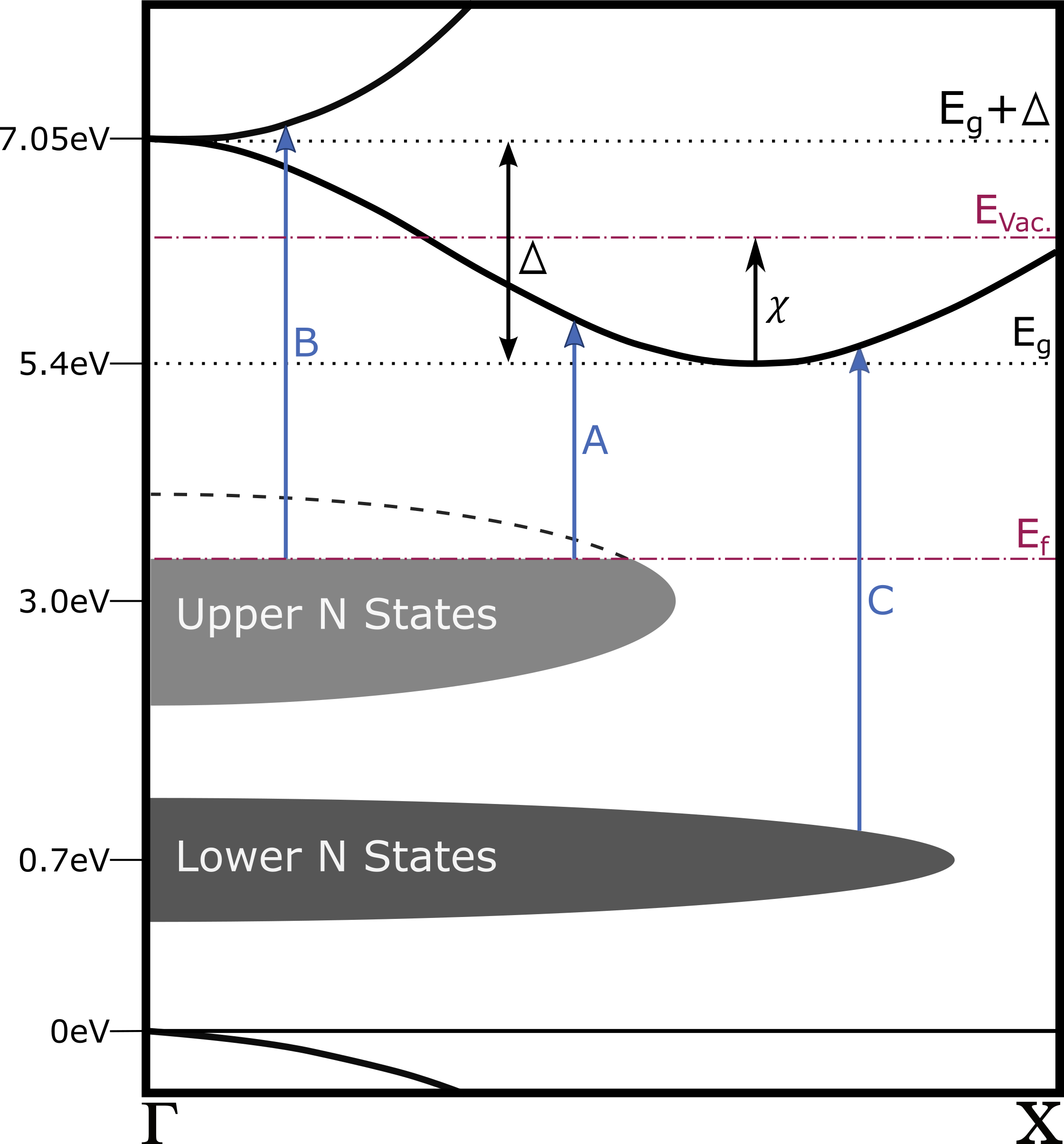}
	   \caption{\label{F1} Band structure of diamond for the $\Gamma\rightarrow$X photocathode emission direction showing the photoexcitation transitions from the upper and lower nitrogen dopant states to the upper and lower conduction bands of diamond accessed by the tunable UV radiation source (see text): E$_g$ is the indirect band gap, E$_g$+$\Delta$ is the direct band gap at the $\Gamma$ point, E$_{vac.}$ is the vacuum energy, $\chi$ is the electron affinity, E$_f$ is the Fermi energy, and the upper and lower nitrogen dopant states are shown centered at 2.4 and 4.7eV below the CBM respectively.}
 \end{figure}

	Figure 1 schematically depicts the positions of these two nitrogen states in the $\Gamma\rightarrow$ X band structure of diamond relevant for the investigated electron emission from the crystal’s (001) face.  Due to its higher photoionization energy (binding energy) and hence a smaller spatial extent of its wavefunction, the rms momentum width of the lower state is shown to be larger than that of the upper state.  Also shown in Figure 1 are the three optical transitions of relevance for these studies that excite electrons into the conduction band of the diamond photocathode; (i) excitation into the lower conduction band (side valleys of diamond) from the upper nitrogen state (labeled transition A), (ii) excitation from the upper nitrogen state into the upper conduction band near the $\Gamma$ point of the Brillouin zone (labeled transition B), and (iii) excitation into the lower conduction band from the lower nitrogen state (labeled transition C).  As the energy difference $\Delta$ between the CBM and the minimum of the upper diamond conduction band at the $\Gamma$ point is 1.65eV, based on measurements of the 5.4eV indirect band gap  \cite{44} and the 7.05eV direct $\Gamma$ point gap  \cite{44,51} at 300K, the onset of transition B will be around $\hbar\omega\approx$4eV ($\lambda\approx$300nm) whereas that for transition A should be in the yellow to blue spectral region ($\lambda\approx$400-600nm).  However, we also expect the cross-section for transition A to be significantly less than that for either transitions B or C.  This is because the overlap integral in k-space for transition A will be less than that of transition B due primarily to the upper nitrogen state’s small size (i.e., variance) in momentum space about the $\Gamma$ point of the Brillouin zone: The matrix element for transition B is consequently larger than that for transition A since the upper conduction band is also centered about the $\Gamma$ point whereas the six lower conduction band minima with their small 0.22$m_0$ effective mass transverse to the $\Gamma$-X direction are located $\sim$75\% towards the X point  \cite{35}.  On the other hand, transition C with an onset at around $\hbar\omega\approx$4.7eV ($\lambda\approx$260nm) is expected to be the strongest as the lower nitrogen state has a much ($\approx$10 times) higher density of states than the upper nitrogen state at our doping level  \cite{36}.  We note that the stated spectral positions of these three absorption bands and their relative strengths is also consistent with prior scientific literature  \cite{52}.

	These three photoexcitation transitions and their relative strengths play the initial key role in determining the spectral photoemission properties of the nitrogen-doped diamond(001) photocathode, primarily because emission of electrons from the upper conduction band is from states with negative electron affinity ($\chi<$ 0) whereas that from the lower conduction band is predominantly from states with positive electron affinity ($\chi>$ 0).
\subsection{Back-side (substrate face) emission}
	The variation of the measured MTE with incident photon energy $\hbar\omega$ for electron emission from the nitrogen-doped substrate is shown by the open circles in Figure 2.  The MTE increases linearly with the incident photon energy up to $\hbar\omega\approx$4.8eV with an onset in the observable photoemitted electron signal at around $\hbar\omega$ = 3.3eV – the photon energy at which the QE is above our $\sim$10$^{-9}$ measurement floor.  The intercept (threshold photon energy) for this linear dependence (solid black line in Figure 2) occurs at $\hbar\omega\approx$ 2.4($\pm$0.1)eV which is in agreement with threshold energy required for internal photoionization of the upper nitrogen state  \cite{36,37} associated with transition A in Figure 1.  More specifically, since the upper nitrogen state is distributed over a band with finite energy width for the 100-200ppm N concentration, this threshold photon energy is reflective of the average energy that the Fermi level resides below the CBM within the substrate region.  Consequently, for this photocathode orientation, where the applied field in the diamond(001) crystal associated with the DC gun’s 1MV/m acceleration field is in opposition to the surface depletion field at the emission face of the $n$-doped substrate, the close agreement between the measured threshold photon energy and the upper nitrogen dopant state donor ionization energy suggests that approximately half of the upper nitrogen dopant states (i.e., a small fraction of the total 100-200ppm dopant level) are ionized in the depletion region of the substrate.  The upward band bending due to the stronger surface depletion field will nonetheless increase the potential for electron emission from the photocathode face \cite{45,49} to a value approaching 1eV.
\begin{figure}
	   \includegraphics[width=\linewidth]{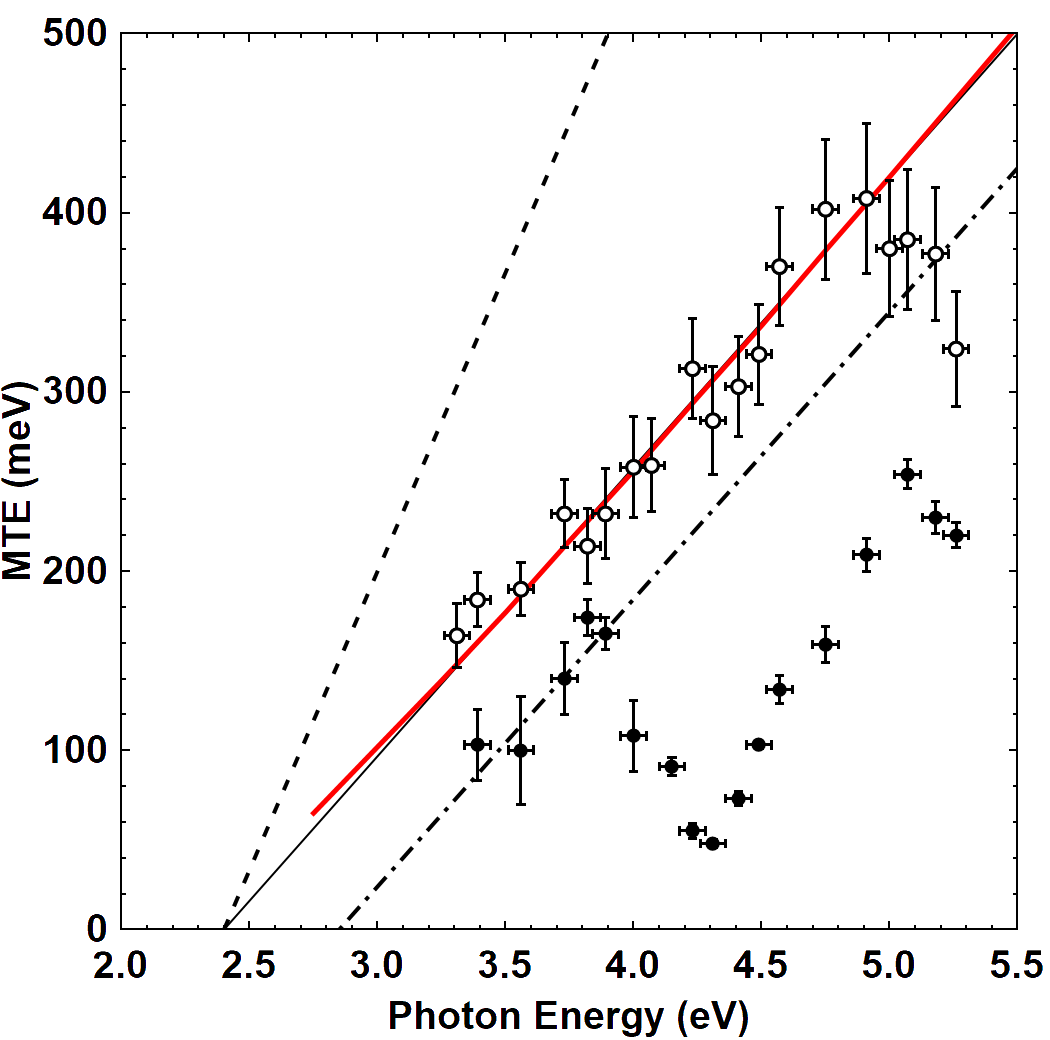}
	   \caption{\label{F2} Mean transverse energy (MTE) of emitted electrons from the nitrogen-doped diamond(001) photocathode as a function of incident photon energy: Experimental measurements for front-side illumination (black data points) and back-side illumination (open circles); a linear fit (black line) to the back-side illuminated MTE variation below $\hbar\omega$= 4.75eV indicating an internal ionization energy $\xi$= 2.4($\pm$0.1)eV; MTE = $\Delta$E/3 using a 2.4eV effective work function (dashed black line); and the linear dependence of the front-side illuminated MTE below $\hbar\omega$= 3.8eV (dot-dashed black line).  The result of a theoretical evaluation of emission from a thermalized electron distribution (energy equipartitioned with the optical phonons) in the upper conduction band of diamond using $\xi$= 2.3eV is shown by the solid red line (see text).  }
 \end{figure}

	It is also notable that the slope of the spectral MTE increase is approximately half of the Dowell-Schmerge dependence \cite{28}, $MTE=\frac{1}{3}\Delta E$  (black dashed line in Figure 2); in this case, the fit shown by the solid black line in Figure 2 is given by $MTE=0.161(\hbar\omega-\xi)$ where $\xi$ = 2.4eV is the threshold ionization energy from above.  This observation can be explained by (i) the uniquely strong electron-phonon coupling in diamond  \cite{40} and (ii) physical characteristics of the emitting diamond conduction band states for an increased $\sim$1eV value of the electron affinity $\chi$. Firstly, as shown in Appendix A within the Boltzmann approximation to the Fermi-Dirac distribution (i.e. a low photoexcited carrier density), an instantaneous thermalization of the electrons photoexcited into either the lower or upper conduction bands of diamond from a defect (or donor) state at an energy $\xi$ below the CBM results in an electron distribution with a thermal energy described by $k_BT_e=\frac{2}{3}(\hbar\omega -\xi)$.  If this initial electron energy is significantly greater that the $\hbar\Omega\approx$160meV optical phonon energy  \cite{31,41}, the strong electron-phonon coupling in diamond  \cite{40} will result in a rapid equipartition of energy between the electron and phonon distributions (see Appendix B) halving the electron energy to $k_BT_e=\frac{1}{3}(\hbar\omega -\xi)$.  We note that one expects electron-phonon coupling to dominate over carrier-carrier scattering as the generated photoexcited carrier densities are very low.  Therefore, the rapid ‘thermalization’ over all available k-space conduction band states to the equilibrium electron energy is primarily by the electron-phonon interaction which will, of course, result in the same final electron energy.

	Secondly, not all conduction band states with positive electron momentum in the $\Gamma\rightarrow$ X direction can efficiently emit their hot electrons from the nitrogen-doped diamond(001) photocathode face.  The discussed increase in the effective electron affinity $\chi$ (potential barrier) to $\sim$1eV due to the surface band upward bending will substantially reduce the emission efficiency from the lower conduction band states for $k_BT_e$ less than $\chi$  \cite{45,50}.  Moreover, electron states in this band between the $\Gamma$ point and the CBM have a negative group velocity (i.e., $\frac{\partial E}{\partial p} < 0$) with respect to the emission direction, implying a much-reduced emission efficiency for electrons in these states as their wave packets are moving away from the emission face.  Consequently, it is likely that only electrons in the thermalized distribution that populate the upper conduction band around the $\Gamma$ point of the Brillouin zone can be efficiently emitted as this band has a NEA of perhaps 0.5eV (or greater).  Based on the recent band structure evaluation of Loffas et al.  \cite{35} and our calculations using the \textsc{QUANTUMESPRESSO} suite  \cite{53}, this band has a relatively isotropic dispersion around the $\Gamma$ point with an effective mass of $m^*$ = 0.4-0.45$m_0$.  As a result, and in accordance with the theory of sub-threshold photoemission from the thermal tail of an electron distribution previously developed for metal photocathodes [19] with $m_0$ modified by $m^*$, the spectral variation of the MTE is expected to be of the form $MTE=(\frac{m^*}{m_0})k_BT_e\approx0.14(\hbar\omega-\xi)$, which is within 15\% of the observed trend (Figure 2).

	A more exact simulation of the photoemission using the analysis presented in Ref. 25, which includes the density of both the emitting band and recipient vacuum states, produces the red line in Figure 2 using a purely parabolic dispersion for the upper conduction band with $m^*$ = 0.43$m^0$.  This simulation fit to the experimental data assumes immediate equipartition of the electron energy with the optical phonons to generate a Boltzmann electron distribution with an energy $k_BT_e=\frac{1}{3}(\hbar\omega -\xi)$ and uses $\xi$ = 2.3eV.  It also employs a positive electron affinity of $\chi$ = 1eV for the doped diamond(001) face which gives an effective electron affinity of -0.65eV for the emitting upper conduction band.  We note that the simulated theoretical MTE values are not strongly dependent on the value of $\chi$ provided that the electron affinity of the upper conduction band is less than about -0.3eV.

	The above interpretation is also consistent with the observed spectral trend in the QE (evaluated as emitted electrons per incident photon) for this back-side illumination case shown in Figure 3.  Specifically, in accordance with theoretical predictions that $QE = A(\hbar\omega -\phi)^2$ where A is a constant  \cite{54,55}, the displayed plot of $(QE)^{1/2}$ against incident photon energy clearly indicates the onset of an additional stronger emission (i.e. absorption) mechanism at $\hbar\omega\approx$4.0eV that does not influence the spectral MTE trend (Figure 2).  As shown by the analysis presented in Appendix A, this is possible for emission from a thermalized electron distribution in diamond irrespective of whether the electrons are photoexcited into the upper or lower conduction band, provided they originate from the same defect (or dopant) state at a fixed energy $\xi$ below the CBM.  For the upper nitrogen state, the initial absorption below $\hbar\omega\approx$4eV (but above $\xi$ = 2.4eV) is then into the six conduction band side valleys (transition A in Figure 1), with the knee point in Figure 3 at $\hbar\omega\approx$4eV being associated with the onset of stronger absorption directly into the upper conduction band (transition B in Figure 1) around the $\gamma$ point of the Brillouin zone.
\begin{figure}
	   \includegraphics[width=\linewidth]{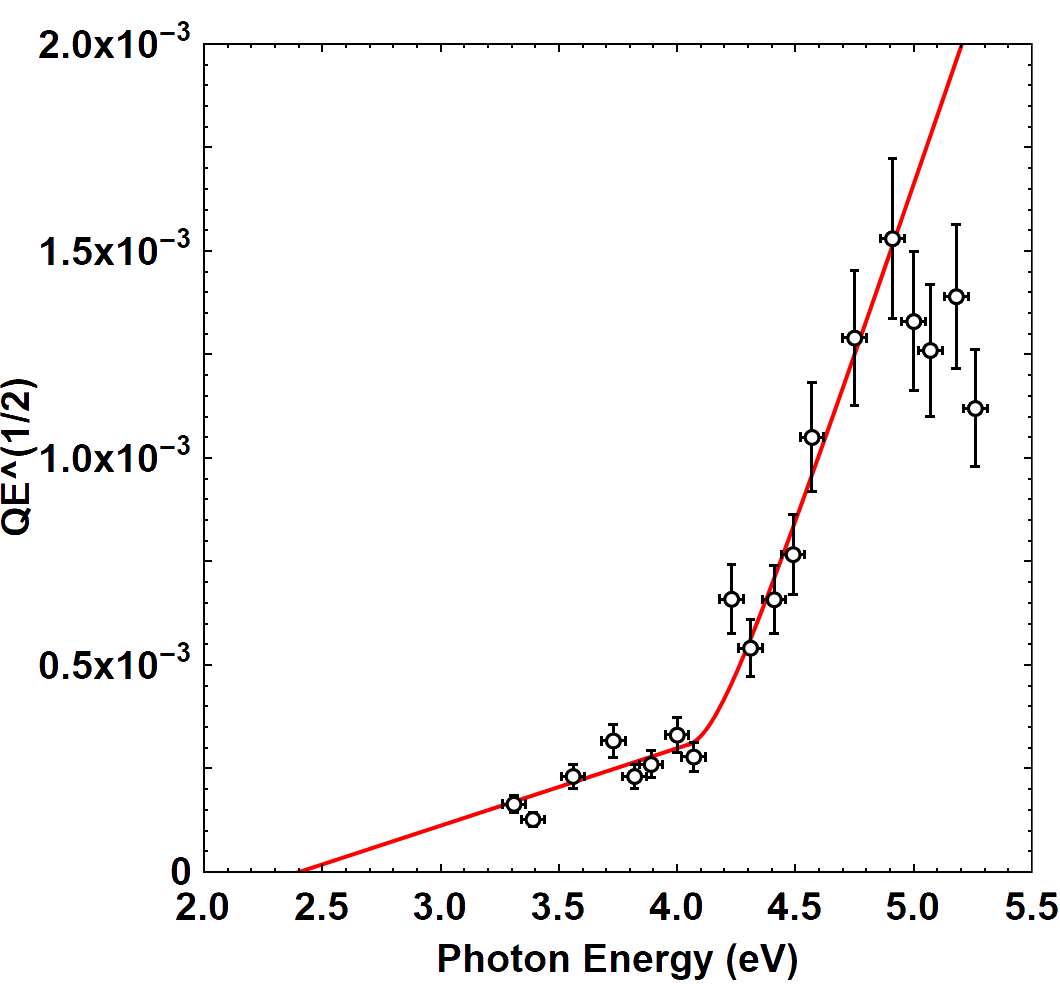}
	   \caption{\label{F3} Square root of the measured emitted electron per incident photon quantum efficiency (QE) as a function of incident photon energy for the back-side illumination of the nitrogen-doped diamond(001) photocathode. A fit to the experimental data below $\hbar\omega\approx$4.7eV assuming two contributions with a quadratic excess energy dependence is shown by the red line; that is $QE=A_1(\hbar\omega-\xi_1)^2+A_2(\hbar\omega-\xi_2)^2$ with $A_1$=3.5x10$^{-8}$eV$^{-2}$, $\xi_1$=2.4eV, $A_2$=2.8x10$^{-6}$eV$^{-2}$, and $\xi_2$=4.05eV.}
 \end{figure}

	This interpretation is supported by the satisfactory fit to the QE measurements below $\hbar\omega\approx$ 4.7eV (red line in Figure 3) using $QE=A_1(\hbar\omega-\xi_1)^2+A_2(\hbar\omega-\xi_2)^2$ with $\xi_1$ = $\xi$ = 2.4eV and $\xi_2$ = $\xi$ + $\Delta$ = 4.05eV; the latter being the expected internal ionization energy for the upper nitrogen state into the upper conduction band (transition B in Figure 1) since $\Delta$ = 1.65($\pm$0.05)eV  \cite{35,43,44}.  The fit also gives $A_1 = 3.5\times10^{-8}eV^{-2}$ and $A_2 = 2.8\times10^{-6}eV^{-2}$.  As the monotonic spectral trend of the MTE for this back-side illumination case (open circles in Figure 2) indicates that all emission below 4.7eV is predominantly from the upper conduction band, the two$ A_i$ ($i$ = 1, 2) coefficients represent thermalized electrons photoexcited from the upper nitrogen states into the lower ($i$ = 1 and transition A in Figure 1) and upper ($i$ = 2 and transition B in Figure 1) conduction bands respectively.  The factor of $\sim$100 difference between the A coefficients is then clearly consistent with the stated expected strength of the transition A versus that of B. 

	Lastly, above $\hbar\omega\approx$ 4.7eV, the photoexcitation of electrons into the conduction band from the lower nitrogen dopant state begins to strongly influence both the MTE and QE of electron emission.  As this state is $\sim$4.7eV below the CBM [36] and $\Delta$ = 1.65eV, all electrons will be excited into the six side valleys of the conduction band (transition C in Figure 1) for the presented measurements since all our incident photon energies are below 5.3eV.  Thus, as $\xi$ for this state is $\sim$2.3eV larger, the excess band energy for photoexcited electrons generated by this strong absorption mechanism is significantly less than that for those photoexcited into the conduction band states from the upper nitrogen state.  This drives a strong reduction in the temperature $T_e$ of the overall photoexcited thermalized electron distribution which results in the observed decrease in the MTE of the electrons emitted from the upper conduction band state for this back-side illuminated case (open circles in Figure 2).  Interestingly, for this photocathode orientation, the QE also decreases above $\hbar\omega\approx$ 4.7eV (Figure 3), indicating that the resulting reduction in the electron temperature of the population in the emitting upper conduction band states is greater than the increase in the overall photoexcited electron population due to the additional contribution from the lower nitrogen states – an effect that is only possible if the electron affinity of this n-type crystal face is indeed sufficiently large to effectively cut off emission from the lower conduction band states  \cite{45,50}.
\subsection{ Front-side (undoped CVD crystal face) emission}
	The measured spectral electron emission properties from the undoped (front-side) face of the diamond(001) photocathode display features in common with those observed from the substrate (back-side) face, but also show striking differences.  Our interpretation of these experimental results relies on the assumption that the nitrogen dopant photoexcitation physics is substantially the same for front- and back-side illumination, thus producing the similar spectral features in the MTE and QE.  The differences are then interpreted to be caused by a lower electron affinity for the undoped emission face of $\sim$0.5eV  \cite{30,45} and the effect that applied DC gun field has on the position of the Fermi level in the nitrogen doped (substrate) region where all the detected photoelectrons originate.  Moreover, the observed spectral trends in the MTE and QE for front-side emission should also be consistent with the expected electron transport physics across the undoped region of the diamond(001) photocathode – a requirement for this photocathode orientation under our experimental conditions ($\hbar\omega <$ 5.3eV) since all the photoemitted electrons must originate in the nitrogen-doped substrate region of the crystal sample.
\begin{figure}
	   \includegraphics[width=\linewidth]{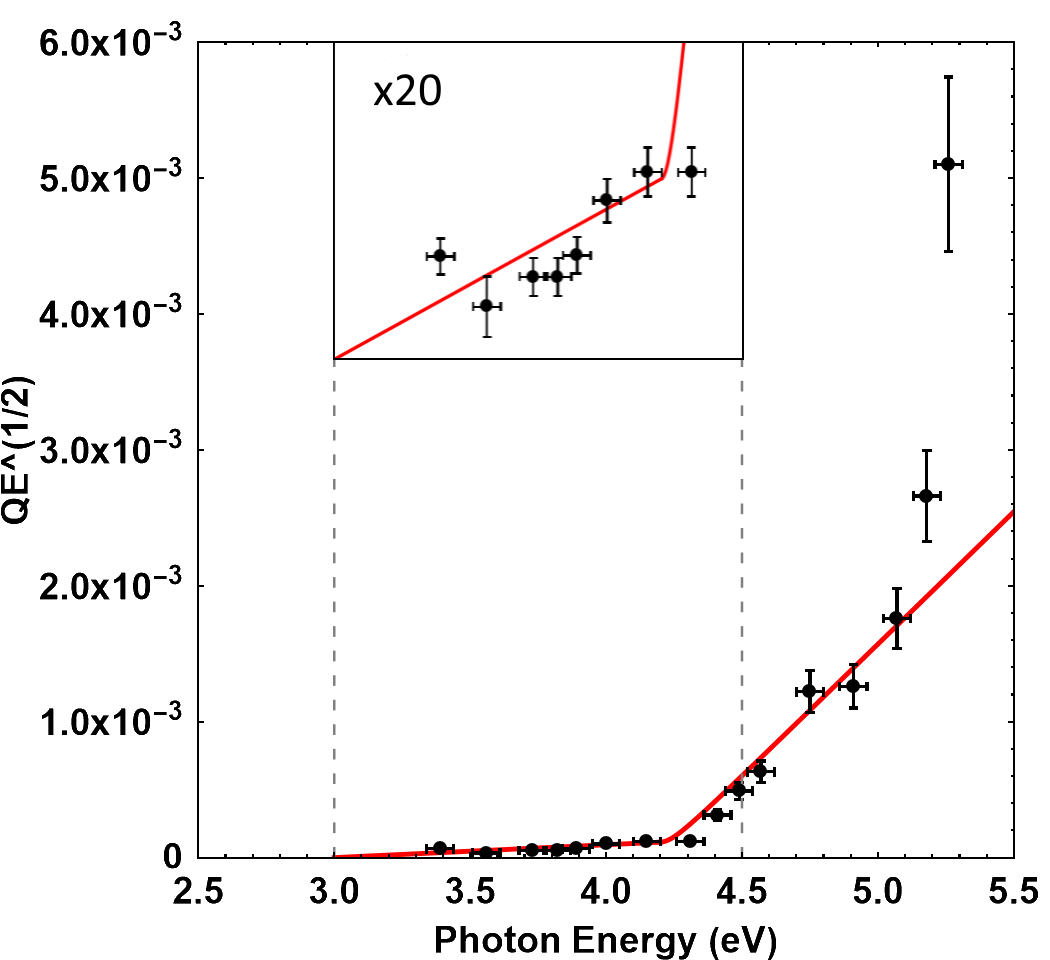}
	   \caption{\label{F4} Square root of the measured emitted electron per incident photon quantum efficiency (QE) as a function of incident photon energy for the front-side illumination of the nitrogen-doped diamond(001) photocathode. A fit to the experimental data below $\hbar\omega\approx$5.0eV assuming two contributions with a quadratic excess energy dependence is shown by the red line; that is $QE=A_1(\hbar\omega-\xi_1)^2+A_2(\hbar\omega-\xi_2)^2$ with $A_1$=9.0x10$^{-9}$eV$^{-2}$, $\xi_1$=3.0eV, $A_2$=3.8x10$^{-6}$eV$^{-2}$, and $\xi_2$=4.2eV. The inset displays the low photon energy dependence in more detail.}
 \end{figure}

	Strong evidence supporting the assumption of essentially the same electron photoexcitation physics comes from the QE for front-side emission (Figure 4) which for $\hbar\omega <$ 4.7eV exhibits a similar spectral trend to that for back-side emission (Figure 3).  Specifically, as for substrate face emission, the spectral QE dependence is again consistent with $QE=A_1(\hbar\omega-\xi_1)^2+A_2(\hbar\omega-\xi_2)^2$  (red line in Figure 4.)  In this case, however, the internal ionization thresholds for transitions A and B (Figure 1) are somewhat larger, the fit giving $\xi_1$ = 3.0($\pm$0.2)eV and $\xi_2$ = 4.2($\pm$0.1)eV, respectively.  This increase is due to the internal applied field (associated with the DC gun) being parallel (rather than in opposition) to the depletion field at the n-type surface for this photocathode orientation.  The large resultant net field pushes the average position of the Fermi level in the doped substrate beneath the peak position of the broad upper nitrogen level located 2.4eV below the CBM  \cite{36}, producing the observed increase in the effective ionization threshold photon energies.  A further consequence of this effect should be that the QE in the photon energy range up to $\sim$4eV is significantly less than that for back-side illumination (Figure 3) – there are simply less populated upper nitrogen states to generate the emitted electrons.  This is indeed the case since the red line fit to the QE data in Figure 4 returns a value for the $A_1$ coefficient of 9$\times$10$^{-9}$eV$^{-2}$ which is 4 times less than that for back-side illumination (Figure 3).

	The initial spectral trend of the MTE below $\hbar\omega$ = 3.8eV observed for emission from the undoped diamond(001) photocathode face (solid black circles in Figure 2) is also very similar to that for substrate face emission; specifically, the same spectral dependence of the form $MTE = 0.161(\hbar\omega - \xi)$ shown by the dot-dashed line in Figure 2 is consistent with the measurements.  The only difference is that the obtained value for $\xi$ is 2.85($\pm$0.1)eV rather than 2.3eV for the doped face emission; but, as required, this value is in agreement with the increased internal upper nitrogen state photoionization threshold $\xi_1$ = 3.0($\pm$0.2)eV extracted from the fit to the spectral QE dependence in Figure 4 for undoped face emission.  Of course, the same internal excess energy dependence exhibited by the MTE for this photocathode orientation below $\hbar\omega$ = 3.8eV has two interconnected implications.  The first is that the initial electron emission is again predominantly from a thermalized electron distribution in the upper conduction band.  And the second is that the band energy of the photoexcited electrons (i.e., their temperature $T_e$ after energy equipartition with the optical phonons (Appendix B)) is essentially unaffected by the required drift transport in the $\Gamma \rightarrow$X emission direction through the CVD-grown undoped diamond layer to its surface under the applied internal field.  The veracity of the second inference, which directly affects the first, requires further justification.

	The second implication can be shown to be quite valid under our experimental conditions for which the $E_{DC}\approx$ 1MV/m acceleration field employed in the DC gun generates an internal field of $E_{DC}/\epsilon_r\approx$ 1.7kV/cm in the photocathode crystal, using a value of $\epsilon_r$ = 5.7 for the relative dielectric permittivity of diamond  \cite{56}.  Under this internal acceleration field, an electron will increase its energy by the $\hbar\Omega\approx$ 160meV optical phonon energy in a characteristic distance $\delta z = \frac{\epsilon_r\hbar\Omega}{qE_{DC}}\approx 1\mu m$, implying that thermalization and energy equipartition of the electrons with the optical phonons will continue through the drift region in the $\sim$0.5mm undoped portion of the crystal.  Moreover, the characteristic time taken by an electron to gain this phonon energy quanta is given by  $\delta t = \frac{\epsilon_r}{qE_{DC}}\sqrt{2m_z\hbar\Omega}$, where $m_z$ is the effective electron mass in the direction parallel to the $\Gamma\rightarrow$ X electron acceleration and emission direction for the diamond(001) photocathode.  For the anisotropic lower conduction band, the longitudinal effective mass $m_z^*\approx$ 1.2$m_0$  \cite{35}, giving $\delta t\approx$ 8.4ps, whereas $\delta t\approx$ 5.0ps for the more isotropic upper conduction band for which $m_z^*\approx$ 0.42$m_0$.  Both characteristic times are close to the 6.7ps optical phonon decay time measured for diamond  \cite{41}.  As a result, during the bulk transport of the electrons to the undoped surface under the applied field, the decay of the optical phonon population is closely balanced by its increase due to energy equipartition with the electrons as they gain energy in the acceleration field.  Thus, to a good approximation, the electron temperature $T_e$ initially generated after photoexcitation and energy equipartition with the optical phonons should be maintained during the transport process.  We also note here that recombination effects are not expected to be significant in the drift transport to the undoped diamond(001) face since the valance band is fully occupied and there are less than 200ppb dopant (or defect) states in the unintentionally doped single-crystal diamond  region.

	Armed with this understanding of the drift transport dynamics and the knowledge of a lower $\sim$0.5eV electron affinity for the undoped diamond(001) face  \cite{30,45} provides the basis for the explanation of the observed striking differences between the spectral MTE dependences of the two studied nitrogen-doped diamond(001) photocathode orientations (Figure 2).  Perhaps the most noticeable difference is non-monotonic spectral dependence of the MTE around $\hbar\omega\approx$ 4eV for emission from the undoped face (solid black circles in Figure 2).  The observed factor of three reduction in the MTE from about 150meV to the minimum value of 50($\pm$10)meV for incident photon energies between 3.8 and 4.2eV can now be understood as being a direct consequence of the lower $\sim$0.5eV electron affinity for the undoped diamond(001) crystal face relative to the  electron energy $k_BT_e$ and the positions of the upper and lower conduction bands.  In general, below $\hbar\omega\approx$ 4.7eV, post transport photoexcited electron emission from the undoped diamond(001) face will have contributions from both the lower and upper conduction band states.  At any given characteristic energy $k_BT_e$ of the thermalized electron distribution, the relative strengths of these two emission contributions can be expected to be primarily dependent upon the value of the surface electron affinity $\chi$ since it directly sets both the positive electron affinity for the lower conduction band and the negative electron affinity of the upper conduction band (i.e., $\Delta - \chi$).  This is because (i) the density of states of the upper conduction band and the one emitting lower conduction band side valley are comparable – the density of states being proportional to $(m_xm_ym_z)^{1/2} \approx 0.25(m_0)^{3/2}$ in both cases – and (ii) the more efficient negative electron affinity emission from the upper conduction band is offset by its lower population density associated with the exponential factor of $e^{(-\Delta/k_BT_e)}$ for the overall non-degenerate Boltzmann distribution (see Appendix A).  Thus, one expects emission from the upper conduction band state to dominate for low electron temperatures ($k_BT_e < \chi$) due to the potential barrier of the positive electron affinity reducing strong emission from the lower conduction band, whereas the opposite is true for higher electron temperatures ($k_BT_e > \chi$) when much more of the thermalized electron distribution in the lower conduction band is above the surface potential barrier for emission into the vacuum.

	This interpretation explains the observed unusual spectral trend in the MTE for electron emission from the undoped diamond(001) face presented in Figure 2 (solid circles).  For the initial lower thermalized electron temperatures below $\hbar\omega\approx$ 3.8eV, emission is mainly from the upper conduction band, giving the same excess internal photoexcitation energy dependence to the MTE as measured for substrate face emission when $\hbar\omega <$ 4.7eV (open circles in Figure 2).  The subsequent remarkable factor of three reduction in the MTE to the minimum value of 50($\pm$10)meV for incident photon energies between 3.8 and 4.2eV can then be associated with the transition from upper to predominantly lower conduction band emission as the thermalized electron energy $k_BT_e$ increases.  The MTE reduction is therefore primarily driven by the decrease in the transverse effective mass of the emitting band states – $m^*\approx$ 0.42$m_0$ for the isotropic upper conduction band and $m^*_T$ = 0.22$m_0$ for the lower conduction band  \cite{35}.  Under the assumption of energy equipartition with the optical phonons, this transition should occur approximately when $k_BT_e\approx\frac{1}{3}(\hbar\omega-\xi)$ from which we estimate that $\chi\approx$ 0.4($\pm$0.1)eV for the undoped photocathode emission face using the experimentally determined value of $\xi\approx$ 2.9eV for the photoionization energy of the upper nitrogen dopant state in this photocathode orientation.  This value of the electron affinity for an undoped diamond(001) face with no surface preparation is consistent with prior work  \cite{30,31,46}.

	The measured minimum value of 50($\pm$10)meV for the MTE at $\hbar\omega\approx$ 4.2eV (filled black circles in Figure 2) is only a little larger than the limiting minimum value of the MTE that can be defined by the lowest electron temperature for which energy equipartition with the optical phonons can exist; that is, when $k_BT_e\approx\hbar\Omega$, so that $MTE\approx\frac{m^*}{m_0}k_BT_e$ = 35meV using 0.22$m_0$ as the transverse effective mass of the lower conduction band  \cite{35}.  This discrepancy is, of course, due to some emission contribution from the upper conduction band with $m^*\approx$ 0.42m0 (half of the emission from each band would give MTE $\geq$ 50meV) and the density of the recipient vacuum states  \cite{25}.  Thereafter, the MTE again increases roughly linearly with photon energy up to $\sim$250meV at $\hbar\omega\approx$ 4.8eV due to the linear dependence of $k_BT_e$ with $\hbar\omega$ (see Appendix A) combined with a concomitant increasing emissive contribution from the upper conduction band (with its larger transverse effective mass $m^*$) as the ratio of electron populations in the upper to lower conduction bands increases with $T_e$.  As a result, the rate of increase of the MTE with incident photon energy for $\hbar\omega\approx$ 4.2 to 4.8eV is greater than that below $\hbar\omega\approx$ 4.7eV for the back-side illuminated case (open circles in Figure 2).  This interpretation of dual emission from the lower and upper conduction bands in the intermediate photon energy range from the undoped diamond(001) face is supported by the relative values of the $A_2$ coefficients extracted using the fits to the measured spectral dependences of the QE from both photocathode faces (red lines in Figures 3 and 4).  Specifically, the value of 3.8$\times$10$^{-6}$eV$^{-2}$ for emission from this undoped face (Figure 4) is 1.5 times larger than that for emission from the doped face (Figure 4), which is consistent with the presence of the additional emission from the lower conduction band.

	Above $\hbar\omega\approx$ 5eV, the MTE for electrons emitted from the undoped diamond(001) face of the photocathode exhibits a further distinct downturn caused, as for substrate face emission, by the strong photoexcitation of electrons with less excess band energy into the lower conduction band from the $\sim$10$\times$ more populous lower nitrogen state (transition C in Figure 1.)  The energetic onset of this MTE decrease is again delayed by $\sim$0.2eV compared to that for substrate face emission at $\hbar\omega\approx$ 4.8eV due to the stronger net internal field in in the doped region for this photocathode orientation.  However, a further distinct difference is that in this case the QE dramatically increases for $\hbar\omega >$ 5eV (Figure 4) rather than decreasing as observed for substrate face emission (Figure 3) above 4.8eV.  This is clearly consistent with a 2-3 times lower electron affinity for the undoped face.  Specifically, the evaluated $\sim$0.4eV electron affinity is sufficiently low to provide for increasing QE while $T_e$ is reduced by the excitation of lower excess energy electrons from the lower nitrogen state which drives the observed reduction in the MTE (filled black circles in Figure 2).  Based on the prior MTE reduction at $\hbar\omega\approx$ 4.0eV and the extrapolation of the QE using a further $A_3(\hbar\omega-\xi_3)$ dependence, the expectation would be that a minimum MTE of around 100meV could be attained at $\hbar\omega\approx$ 5.5eV with perhaps a QE  $\sim$10$^{-4}$ for front-side illumination of this nitrogen-doped single-crystal diamond(001) photocathode.  As this photon energy is above the 5.4eV indirect band gap [44], phonon-assisted band-to-band absorption into the CBM is also possible which could further enhance the QE.  Moreover, this may occur in combination with an additional reduction in the MTE, due to the photoexcitation of more colder electrons into the CBM, thereby aiding the ‘transverse momentum filtering’ of the emitted electrons that is facilitated by the low 0.22$m_0$ transverse effective mass of the lower conduction band  \cite{35}. 
\section{Summary}
	The presented spectral characterization of the emission properties from both the undoped and nitrogen-doped faces of a diamond(001) photocathode are found to be fully consistent with the well-established photoexcitation physics  \cite{36,37}, electron thermalization (Appendices A and B) and strong optical phonon scattering dynamics  \cite{40,41}, conduction band dispersions  \cite{35}, and effective surface photoemission barriers (i.e., electron affinity and band bending)  \cite{30,32,45,50} of diamond.  In particular, the experimental measurements provide clear evidence for the restriction on the MTE that is placed by the transverse effective mass $m_T^*$ of the emitting band states for the observed predominantly sub-threshold photoexcited thermionic emission of hot electrons at temperature $T_e$ from diamond’s conduction band; that is, $MTE\approx(\frac{m^*_T}{m_0})k_BT_e$, for $m^*_T < m_0$.  For emission from the doped face with the higher effective electron affinity, the QE and MTE measurements also verify that the temperature of the emitting thermalized electron distribution from the upper conduction band (which, in this case, is also energy equipartitioned with the optical phonons) is only dependent upon the energy difference between the incident photon energy $\hbar\omega$ and the internal ionization energy $\xi$ of the dopant (or defect) state from which the electrons originate even if the electrons are photoexcited into different bands (see Appendix A).  In the present case of diamond, with its strong electron-(optical)phonon interaction  \cite{40} that leads to energy equipartition between the electron and optical phonon distributions (see Appendix B), this then implies $MTE \approx \frac{1}{3}(\frac{m^*_T}{m_0})(\hbar\omega-\xi)$ – an expression that is consistent with the presented experimental data for $\Gamma\rightarrow$ X emission.  Further, spectral regions are observed for which the MTE displays an anomalous dependence (decreasing with increasing incident photon energy) and are shown to be due to (i) the injection of colder electrons into the conduction bands from the lower nitrogen state when $\hbar\omega\approx$ 4.7eV and (ii) increased relative emission from the lower conduction band (with its lower transverse effective mass) when $k_BT_e$ becomes greater than the electron affinity $\chi$ as deduced for the case of emission from the undoped diamond(001) face.

	Although emission from the Boltzmann tail of a thermalized electron distribution photoexcited sub-picosecond timescales has already been observed in the semiconductors GaSb(001) and InSb(001)  \cite{57} and more recently in a Cu(111) photocathode  \cite{58}, emission from the nitrogen-doped diamond(001) photocathode differs in two important ways.  First, analysis of the experimental data indicates that hot electron thermionic emission has been observed from a band into which electrons were not initially photoexcited.  Similar effects are expected in other materials on sub-picosecond timescales, particularly semiconductor photocathodes like Cs$_2$Te and Cs$_3$Sb that possess a conduction band below the band states nearer the vacuum level that emit the photoexcited electrons  \cite{59}.  Second, in a further commonality with semiconductor photocathodes  \cite{1}, the electron emission is not on ‘prompt’ sub-100fs timescales since, for both orientations of the photocathode, electrons must drift to the emitting surface; for the front-side illuminate case through at least the $\sim$0.5mm homoepitaxially grown undoped diamond crystal and for back-side illumination on average half of the $\sim$0.4mm nitrogen-doped substrate thickness.  We estimate that this electron drift time in the applied DC field is a few nanoseconds, which is much greater than the $\sim$2ps propagation time for the incident $\sim$0.5ps UV excitation pulse through the substrate region where the photoelectrons originate.

	The MTE measurements for the front-side illuminated case are also of potential practical interest as they indicate that band transport through a suitable transparent surface layer material with a low transverse effective mass could be used as a transverse ‘momentum filter’ to reduce the intrinsic emittance of solid-state photocathodes.  Such a photocathode design would require careful engineering of the interface between the electron source material and the surface layer crystal to assure good electron injection efficiency and hence an acceptable overall QE.  We note that coating deposition has already been successfully employed to protect a spin-polarized photocathode  \cite{60} and that ohmic contacts between some wide-bandgap materials and metals are available, including for diamond  \cite{61}.
\appendix
\section{Appendix I}
	The following analysis shows that under the assumption of rapid thermalization of photoexcited electrons into a Boltzmann distribution (i.e., for a population density well removed from degeneracy) extending over two conduction bands the temperature $T_e$ of the electron distribution is only dependent upon the excess photoexcitation energy above the bottom of the lower conduction band (the conduction band minimum (CBM)).  Further, this result is not dependent on whether the electrons are photoexcited into the upper or lower conduction band, provided that they originate from the same defect (or dopant) state at a fixed energy $\xi$ below the CBM. 

	For simplicity, the two conduction bands, separated by an energy $\Delta$, are considered to be symmetric with parabolic dispersions characterized by effective masses of $m_1$ and $m_2$ for the lower Band 1 and the upper Band 2 respectively.  If electrons are photoexcited from the defect (or dopant) state at an energy $\xi$ below the CBM of the lower conduction band with photons of energy $\hbar\omega$ between $\xi$ and $\xi+\Delta$, then they are excited only into Band 1 with an excess energy above its CBM of $\hbar\omega-\xi$.  After thermalization over both bands to an electron temperature $T_e$, the number density of electrons in each band $n_i$ (i = 1,2) is readily determined;
\begin{subequations}
\begin{equation} n_1 = Ag_1\int_0^{\infty}dE\sqrt{E}e^{(\frac{-E}{k_BT_e})}=\frac{1}{2}Ag_1\sqrt{\pi(k_BT_e)^3}\end{equation}
\begin{equation}\begin{split}n_2 & = Ag_2\int_\Delta^\infty dE \sqrt{E-\Delta} e^{(\frac{-E}{k_BT_e})} \\ & =\frac{1}{2}Ag_2\sqrt{\pi(k_BT_e)^3}e^{(\frac{-\Delta}{k_BT_e})}\end{split}\end{equation}
\end{subequations}
	Here $A = \sqrt{2}/(\pi^2\hbar^2)$ is a constant, the $g_i$ parameters are equal to $(m_i)^{3/2}$ and so reflect the effective mass dependence of the band density of states, and the zero of energy is set to the bottom of the lower conduction band so that the states of upper conduction band (Band 2) start at an energy $\Delta$.  The average energy $E_i$ of the electrons in each band can then be evaluated;
\begin{subequations}
\begin{equation}\begin{split} n_1E_1 & =Ag_1\int_0^{\infty}dE.E\sqrt{E}e^{(\frac{-E}{k_BT_e})} \\ & =\frac{3}{4}Ag_1\sqrt{\pi(k_BT_e)^5} \\ & =(\frac{3}{2}k_BT_e)n_1\end{split}\end{equation}
\begin{equation}\begin{split}n_2E_2 & = Ag_2\int_\Delta^\infty dE.E\sqrt{E-\Delta}e^{(\frac{-E}{k_BT_e})} \\ & =n_2\Delta+\frac{3}{4}Ag_2\sqrt{\pi(k_BT_e)^5}e^{(\frac{-\Delta}{k_BT_e})} \\ & =(\Delta+\frac{3}{2}k_BT_e)n_2\end{split}\end{equation}
\end{subequations}
	The sum of these two expressions must equal the initial excess photoexcitation energy $\hbar\omega-\xi$ multiplied by the total number of photoexcited electrons $n_1+n_2$; that is, 
\begin{subequations}
\begin{equation}(\hbar\omega-\xi)(n_1+n_2)=(\frac{3}{2}k_BT_e)n_1+(\Delta+\frac{3}{2}k_BT_e)n_2\end{equation}
\begin{equation}\Rightarrow(1+\frac{g_2}{g_1}e^{(\frac{-\Delta}{k_BT_e})})(\Delta+\frac{3}{2}k_BT_e-\hbar\omega+\xi)=\Delta\end{equation}
\end{subequations}
using $\frac{n_1}{n_2}=\frac{g_2}{g_1}e^{(\frac{-\Delta}{k_BT_e})}$ from equations (A1).
	The transcendental equation (A3b) yields the temperature for the thermalized electron distribution and must, in general, be solved numerically.  However, for our case of photoexcitation of electrons from nitrogen dopant states into the conduction bands of diamond, the exponential factor $\frac{g_2}{g_1}e^{\frac{-\Delta}{k_BT_e}}$ can be safely neglected.  This is because (i) $g_1$ = 6($m_xm_ym_z$)$^{1/2} \approx$ 1.45($m_0$)$^{3/2}$ for the six side valleys of the diamond conduction band for which $m_z$ = 1.2$m_0$ and $m_{x,y}$ = 0.22$m_0$  \cite{35}, (ii) $g_2$ = 0.27($m_0$)$^{3/2}$ as the upper conduction band has an effective mass of 0.42$m_0$, and (iii) $\Delta > k_BT_e$ for all our experimental conditions after thermalization and energy equipartition with the optical phonons (see Appendix B).  As a result,  $\frac{g_2}{g_1}e^{\frac{-\Delta}{k_BT_e}}\leq$ 0.05 for all the experimental conditions described in this paper, implying that the thermalized population density in the upper conduction band of diamond is much less than that in the six side valleys of the lower conduction band; that is, $\frac{n_2}{n_1}\ll1$.  Within this approximation, equation (A3b) yields $\frac{3}{2}k_BT_e\approx\hbar\omega-\xi$ which after energy equipartition with the optical phonons (Appendix B) gives the relation used in this paper;
\begin{equation}k_BT_e\approx\frac{1}{3}(\hbar\omega-\xi)\end{equation}
	For the case of electron photoexcitation only into the upper conduction band from a defect (or dopant) state, as can clearly happen in nitrogen doped diamond (Figure 1), the above analysis with the same approximation and energy equipartition again yields equation (A4) since the initial excess photoexcitation energy is again $\hbar\omega-\xi$ with the same zero energy at the bottom of the lower conduction band (Band 1) and thus equations (A1) and (A2) apply.
\section{Appendix II}
	Following the photoexcitation of hot electrons into the conduction bands of diamond, the electrons will promptly, within a few femtoseconds  \cite{40}, emit optical phonons of energy $\hbar\Omega\approx$ 160meV  \cite{32,39} due to their strong mutual interaction.  This will lead to a rapid build-up of the optical phonon population producing a ‘phonon bottleneck’ for the cooling of the electron distribution.  In diamond, the long optical phonon lifetime of $\sim$7ps  \cite{41} will then allow an equilibrium to be established between the temperatures of the electron and optical phonon populations – an equipartition of the energy between the two distributions.  Ignoring spin, as the electron-phonon interaction is spin independent, and noting that the experimental electron population photoexcited into the upper conduction bands is highly non-degenerate, the electrons can be considered to have three translational degrees of freedom.  For the case considered in this paper, conduction band electron-optical phonon scattering in diamond, the optical phonons can also be considered to have three degrees of freedom; (i) the longitudinal (LO) and transverse (TO) optical phonons are degenerate at the $\Gamma$ point  \cite{40} giving three dimensional degrees of freedom for scattering with the electrons in the upper conduction band centered on the $\Gamma$ point of the Brillouin zone, and (ii) the optical phonons interacting with the electrons in the six lower conduction band side valleys along the $\Gamma$ - X direction will be predominantly longitudinal  \cite{40} again giving effectively three degrees of freedom.  Further, in common with the non-degenerate hot electron population photoexcited in diamond from the Nitrogen dopant states, the generated optical phonon population may be described using a Boltzmann distribution since, for the lattice temperature $T_L$ = 300K in the measurements, $k_BT_L$ is much less than the $\hbar\omega\approx$ 160meV phonon energy.  As a result, equipartition of the energy between the electron and optical photon distributions implies that half of the initial photoexcited electron energy $\hbar\omega-\xi$, for electrons excited from a state at an energy $\xi$ below the CBM, will be lost to the optical phonons in less than 10fs; that is, after energy equipartition and thermalization, we have
\begin{equation}\frac{3}{2}k_BT_e\approx\frac{1}{2}(\hbar\omega-\xi)\end{equation}
giving equation (A4) (see appendix A.)
\begin{acknowledgments}
This work was supported by the U.S. Department of Energy under award number DE-SC0020387.  The authors gratefully acknowledge discussions with C. Grein, T. Vecchione, and B. Dunham and the assistance of the Liberal Arts and Sciences Machine Shop at the University of Illinois at Chicago.  The authors also thank Ir-Jene Shan for his assistance with the preparation of this manuscript.
\end{acknowledgments}

\bibliography{NDiamond}

\end{document}